\documentclass{aadebug}

\corr{0309027}{225}

\begin{document}


\runningheads{Bil Lewis}{Debugging Backwards in Time}

\title{Debugging Backwards in Time\\
}

\author{
Bil~Lewis\addressnum{1}
}

\address{1}{
Lambda Computer Science,
352 Central Ave,
Menlo Park, CA, 94301, 
USA
}

\extra{1}{E-mail: \{Bil.Lewis\}@LambdaCS.com}


\begin{abstract}
 
By recording every state change in the run of a program, it is
possible to present the programmer every bit of information that might be
desired. Essentially, it becomes possible to debug the program by
going ``backwards in time,'' vastly simplifying the process of
debugging. An implementation of this idea, the ``Omniscient
Debugger,'' is used to demonstrate its viability and has been used
successfully on a number of large programs. Integration with an event
analysis engine for searching and control is presented. Several
small-scale user studies provide encouraging results. Finally
performance issues and implementation are discussed along with
possible optimizations.

This paper makes three contributions of interest: the concept and
technique of ``going backwards in time,'' the GUI which presents a
global view of the program state and has a formal notion of ``navigation
through time,'' and the integration with an event analyzer.

\end{abstract}


\section{Introduction}

Over the past forty years there has been little change in way commercial
program debuggers work. In 1961 a debugger called ``DDT'' \cite{DDT:1962}
existed on Digital
machines which allowed the programmer to examine, deposit, set break points,
set trace points, single step, etc. In 2003 the primary debuggers for Java
allows one to perform the identical functions with greater ease, but little
more.

I believe that the reason for this is that the designers of debuggers have
all concentrated on answering one question: ``What information can we provide
to the programmers while the program is running?'' While this is not at all
unreasonable, it's not the right question.  The question they should have
been asking is ``What information will help the programmer most?'' The answers
are not the same.

\paragraph{Omniscient Debugging}

An omniscient debugger works by collecting events at every state change
(every variable assignment of any type) and every method call in a
program. After the program is finished, it brings up a debugger display and
allows the programmer to look at the state of the program at any time
desired. The programmer can select any variable and go ``backwards in time''
to see where it was set, or what its values were. It is possible to ``single
step'' the program forwards or backwards, to step to any method call, follow
any exception throw, any context switch, etc.

The objective of this paper is to introduce the concept of omniscient
debugging and to illustrate by use of a specific implementation that it is
effective. The details of implementation and performance are interesting 
only as they form a lower bound. If it proves to be effective, there are
lots of ways of improving performance. 

\paragraph{Omniscient Debugging is Interesting}

There are a number of things that make omniscient debugging worth the effort of exploring. 

\begin{itemize}

\item First and foremost, (I claim) debugging is easier if you can go backwards. The most common
question programmers have is ``Who set this variable?''

\item It eliminates the worst problems with breakpoint
debuggers: no ``guessing'' where to put breakpoints, no ``extra steps'' to
debugging, no ``Whoops, I went too far'', no non-deterministic problems.  The
programmer will never have to run the program twice.

\item It gives the programmer a unique view of the program. Being able to
see the traces of all the method calls for each of the threads is quite
simply interesting. It makes looking at unknown code much easier.

\item All the data is serializable. It can be analyzed remotely. It can be
saved to a file. Several programmers can work with the exact same debugging
session.  Beta customers can e-mail a debugging session to developers even
if they don't have the source.

\end{itemize}

Section 2
describes the appearence and functionality of the ODB (an implementation in
Java). Section 3 tells how it is used on different types of bugs and discusses
the results of user studies. Section 4
describes an implementation, its performance and limitations. Section 5
talks about its integration with event analyzers.


\section{The ODB}

For the rest of this paper, we are going to discuss the issues of Omniscient
Debugging within the context of the ``ODB'', an implementation of this concept
in Java. It's important to note that the general concept of Omniscient
Debugging is independent of the implementation language and can equally well
be done in C or C++.

The ODB is an implementation in pure Java which collects information
by instrumenting the byte code of the target program as it's
loaded. It uses a single high-level lock to order the events from
different threads. It has been tested on MacOS, UNIX, and Windows on a
large variety of programs.  It is freely available for download from
\verb+www.LambdaCS.com+. It is a proof-of-concept and serves as an
upper bound for the cost of collection and display.  It is also a
very effective debugger.

For a graphical debugger such as the ODB, there are two vitally important issues we need
to address: presentation of information and ``navigation.'' How does the
programmer get the debugger to display the program state of interest and how
does he recognize that state?

\subsection{Maintaining State}

We want to be able to ``revert'' the program to any previous
state. Effectively this means that every change in every accessible object
or local variable constitutes a separate state and must be recorded. We'll
need to record each assignment in each thread and define an ordering on
them.

The time stamp (an integer) will be the sole referent to program state. Usually it
will not be directly evident. The programmer will select a trace line
to look at, switch to a different thread, move up on down the stack,
or even step through the code. Most of the time he won't even notice
the time stamp, but all display panes will be updated to reflect it.

\begin{figure}[!htb]
\begin{center}
\rotatebox{270}{\includegraphics[width=14cm]{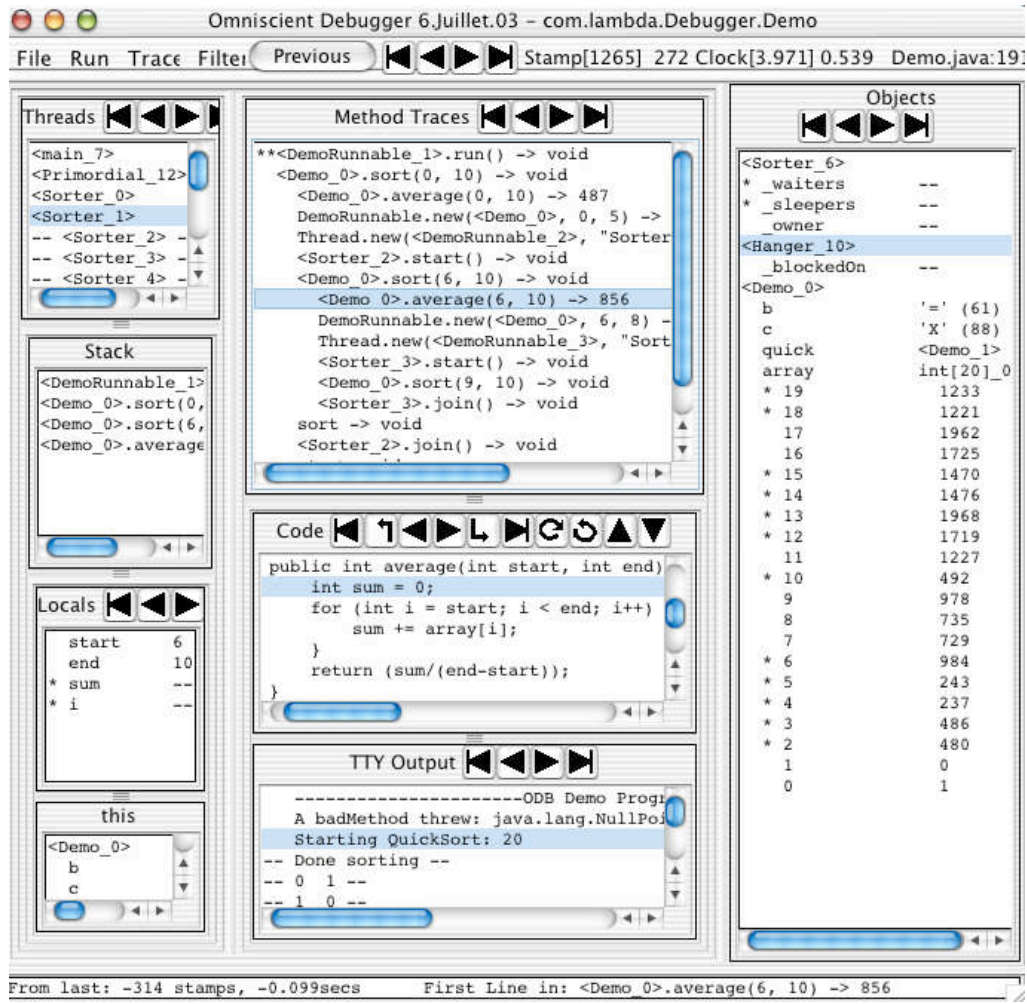}}
\end{center}
\caption{Main Debugger Window}
\end{figure}

\subsection{Presentation}

Every object, variable, I/O stream, etc. will have a known value at each
time stamp. When we change the currently displayed time stamp (we will hence
forth refer to this as ``reverting the debugger'' to a given time) all of the
data displays will be updated to reflect the values at that time.  The line
of code which generated the event will be highlighted, the method trace
that it happened in too. The panes for the current thread, the current stack, the variable
values, etc. will all reflect the selected time stamp.

Things that don't change their value between the previously selected time
stamp and the current one, shouldn't change either. They shouldn't change
their appearance, their position, or their highlighting. Things that do
change should do so obviously. New variable values should be marked, code
lines should be highlighted, etc. The programmer should never wonder ``Did
this change?'' or ``Where am I?''

\subsection{Recognizing State}

It is essential be able to recognize state. This basically means the
programmer should be able to look at an object and know which object
it is and what values its instance variables have.  The ODB supplies a
print format showing the class name and an index number:
\verb+<MyObject_75>+, chopping off the package.  Class objects are
displayed as just the class name (e.g., \verb+Person+), strings and
primitives as expected (\verb+1+, \verb+true+, \verb+"A string"+).

This representation can be easily displayed in a \verb+JList+, or
stored in an editor. It can be printed out and it can even be typed
back in to the debugger at a later date to reference that object. The
object \verb+<Demo_0>+ in figure 1 shows its four instance variables,
the last of which (\verb+array+) is open recursively. In the locals
pane the dashes (\verb+--+)
indicate that a variable hasn't been set yet.

There are a few cases where a different print format is more
``intuitive'' for the programmer. A blocked thread will acquire a
``pseudo variable'' to show the object it's waiting for
(\verb+_blockedOn+) as
will objects whose locks are used.

The collection classes constitute another special case. In reality, a
\verb+Vector+ comprises an array of type \verb+Object+, an internal
index, and some additional bookkeeping variables which are not of any
interest to anyone other than the implementer. So the ODB displays
vectors and hashtables in formats more consistent with the way they
are used. For example, the keys for a hashtable
won't disappear or move.

Large objects are problematic and remain an open issue. What is the best way
to display a 1000x1000x1000 array? Or even just a 10000 element array?  Or a
large bitmap? Or a tree?

\subsection{Displaying Method Traces}

Every method call that is recorded will be displayed in a ``method
trace'' pane. The format of the trace line will be:
\\\verb+<Object>.methodName(arg0, arg1) -> returnValue+. 
\\Each line will
be indented according to depth and a matching ``return line'' will be added
when it returns. Return
lines which directly follow their trace line will be elided.

In figure 1 we can see that current time stamp is no. 272 (out of 1265) and is 
the first event in a method. It is a call to
\verb+average(6, 10)+, which returned 856. This occurred in thread \verb+<Sorter_1>+. From both the
``stack'' pane and the ``trace'' pane 
it's evident that \verb+sort(6, 10)+ is calling
\verb+average()+, 
and that it, in turn, is called recursively from \verb+sort(0, 10)+. The
\verb+this+ object is \verb+<Demo_0>+, which the programmer has copied to the ``objects''
pane and then expanded the \verb+array+ instance variable, which happens to be an
array of 20 integers.

In multithreaded programs, each thread will have its own set of trace lines
which will only be displayed when the debugger is reverted to a time stamp
in that thread. 

Although interesting in itself, the primary purpose of the trace
window is not for debugging directly, but rather as a convenient index
into the time sequence of the program. In other words, the programmer
is going to use the traces to find interesting points in the
program. The programmer will then revert to the time when that method
call was executed and start looking at state.

The ``objects'' pane shows whatever objects the user has copied here
via double-clicking on objects in other windows (any object in any
pane may be copied here). Double-clicking on an instance variable will
expand the value of that instance variable in-place (like the array
\verb+int[20]_0+).

Variables which changed values from the previously selected time stamp will
be displayed with a leading `*' (e.g., \verb+start+ and \verb+end+).


\subsection{Navigation}

Simple navigation through the program's history is accomplished through
selecting lines in a pane or pushing the buttons above them. Selecting a
line in the ``threads'' pane reverts the debugger to the nearest event in that
thread. Selecting a line in the stack frame will display that stack
frame. Selecting a trace line, a code line, or an output line will revert to
the time when that line executed (was printed).

The buttons work as uniformly as possible for the different panes.
The four most common buttons revert to the first, previous, next, and
last time stamp for the selected object in that pane. In the
``threads'' pane, these execute first/last time stamp for the thread,
but previous/next context switch. In the ``code'' pane, they are
first/last line in current method, and ``step into forwards'' and
``step out of backwards.'' The other buttons execute ``step over'',
``return from'', and ``next iteration on current line.''

In addition to first/last/previous/next, it is possible to bring up a list
of all the values a variable ever had and select one of them. The ODB
will revert to the time when that variable was assigned that value.

\begin{figure}[!htb]
\begin{center}
\includegraphics[width=6cm]{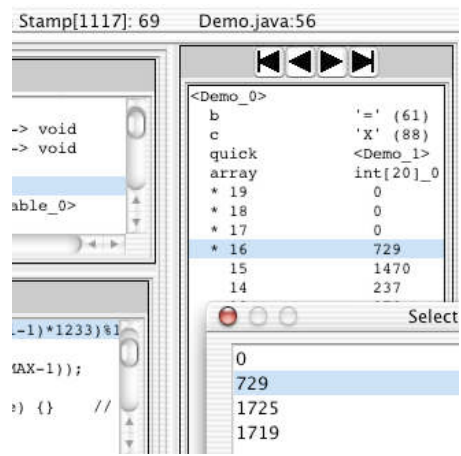}
\end{center}
\caption{Selecting a Value}
\end{figure}

In figure 2 we see a selection from among the four values that array element 16
took on during the run of quick sort.

Modeled after EMACS, the ODB has a minibuffer which is used to present messages and to
execute extended commands (such as \emph{Evaluate Expression} or
\emph{Search}). The programmer may do an incremental text search through the
``trace'' pane, or an event-analysis search through all events, forwards or backwards.

\subsection{Evaluating Expressions Interactively}

Arbitrary method calls may be made at any time. The programmer may
revert the debugger to any time and then evaluate a method call using
the then-current values of the recorded objects.  These methods will be
executed in a different ``timeline'' and won't affect the original
data. In the secondary timeline it is possible to change the value of
instance variables before executing a method.

\section{The Nature of Bugs under the ODB}

The ODB divides bugs along two dimensions:
Do all the events recorded for the bug fit in memory?
Does the bug output wrong data or does it fail to output correct data?

\subsection{The Snake in the Grass}

This second is ``the snake in the grass'' metaphor.  If the program
prints out something wrong (``the answer is 41'' instead of 42), then
we have a handle on the bug. We can see the snake in the grass and we
can grab its tail. Now if the program simply fails to output the correct
answer, then we don't have a direct handle to the bug. This is the
snake in the grass we can't see.

For programs that fit and we can see the snake, then the ODB is absolutely
wonderful. It is possible to start on the faulty output and step backwards,
finding the improper value at each point, and follow that value back to its
cause. (If you've got a snake's tail and you pull on it long enough, you
will get to its head.)  It is not unusual to start the debugger, select the
faulty output, and navigate to its source in 60 seconds. With perfect
confidence.

A good example of this is when I downloaded the JUSTICE java class verifier.  I
found that it disallowed a legal instruction in a place I needed it.
Starting with the output ``Illegal register store'', it was possible to
navigate through a dozen levels to the code that disallowed my instruction.
I changed that code, recompiled, and confirmed success. This took 15 minutes
(most of which was spent confirming the lack of side-effects) in a complex
code base of 100 files I had never seen before.

Breakpoint debuggers suffer from the ``lizard in the grass'' problem. Even when
they can see the lizard and grab its tail, the lizard will break off its tail
and get away. And then they're back to the ``lost lizard in the grass'' problem.

For relatively small problems (say 10,000 events), it is easy just to scan
the entire set of method traces. For larger problems, the ``lost snake in
the grass'' problem becomes more problematic. We have a large data set and
an uncertain idea of what we're looking for.

Happily, searching through large sets of method traces for uncertain
situations is a well-studied problem. The ODB leverages directly off
of this work and incorporates the \verb+get()+ function of Ducass\'e
\cite{DUCASSE:1999opium} as an minibuffer command. Of course now the \verb+get()+
can search backwards as well as forwards.

A good example of this is the quick sort demo that is part
of the ODB jar file. It sorts everything perfectly, except
for elements 0 and 1 sometimes. To search for this problem,
we can issue a \verb+get()+ search to find all calls to
\verb+sort()+ that include 0 and 1.

\begin{verbatim}
port = call & callMethodName = "sort" & arg0 = 0 & arg1 >= 1
\end{verbatim}

There are four matches and \verb+sort(0, 2)+ is the obvious call to look
at. Stepping through that method, it's quickly obvious that \verb+average()+
failed to include the high value, which is why sort \emph{usually} worked.

\subsection{Size}

The other important dimension is that of size. In a 31-bit address
space, there is room to store about 10 million events. A huge percentage
of real bugs fit nicely into this space. A good percentage don't.
There are a number of ways to attack this problem. We can:

\begin{itemize}

\item ``garbage collect'' old events and throw them away.

\item instrument fewer methods.

\item record for a shorter time.

\end{itemize}

\subsubsection{Garbage Collection}
Garbage collection is good because it requires no effort and
effectively maintains a window of events surrounding the bug. It is
bad because it's not throwing away garbage, but just older events
which might be important. It is also bad because it allows the program
to run long enough for performance to become an issue. It adds an
extra 50\% on top of the cost of the average event.

When the source of the bug is within 10 million events of the time
when recording was turned off, then we are back to the well-known
``snake in the grass problem.'' When it is further away, we've lost
the source of the bug, and the garbage collector loses its value.

During the development of the GC, a bug occured sporadically after
the second invocation of the GC. An index into the time stamp array
was getting changed to account for a special situation during
recording, but the GC didn't know it. It would write in a forwarding
index, which would not be recognized as illegal until the second GC,
by which time the original faulty object was long gone.

Attempts to debug this problem with breakpoints and print statements
failed completely. Using the ODB on itself (which was tricky back then)
revealed the problem in relatively short order.

\subsubsection{Safe Code}

It is quite normal for a large percentage of a program to consist of
well-known, ``safe'' code. (These are either recomputable methods, or
methods we just don't want to see the interiors of.)  By requesting
that these methods not be instrumented, a great deal of uninteresting
events can be eliminated.  The ODB allows the programmer to select
arbitrary sets of methods, classes, or packages, which can then be
instrumented or not.

The more work the program does in uninstrumented code, the lower the
overhead.  A good example of this is the following loops:

\vspace{0,5cm}
\begin{tabular}{lr}
Code & Slowdown \\
\verb-for (int i=0; i<MAX; i++) sum+=smallArray[i];- & 300x \\
\verb-for (int i=0; i<MAX; i++) x=x*x+x;- & 100x \\
\verb-for (int i=0; i<MAX; i++) sum+=bigArray[i];- & 30x \\
\verb-for (int i=0; i<MAX; i++) s="Item"+i;- & 2x \\
\end{tabular}
\vspace{0,5cm}

These numbers are very sensitive to memory usage issues. The small
array fits into cache, so the uninstrumented loop runs in three
cycles. The large array (1,000,000 elements) doesn't fit, and
the uninstrumented loop runs 10x slower.

\subsubsection{Starting/Stopping Recording}

If the programmer suspects that a certain known event always occurs
before the bug (e.g., ``When I push this button it crashes.''), then
recording could be turned on at that point and turned off after the
bug. The ODB allows manual control (there's a ``Start/Stop Recording''
button). It also allows automatic control. Automatic control means
that we're going to scan a large set of events for a particular
pattern, which will turn on/off recording. Once again, this is
precisely the problem event analyzers are intended for and the ODB
uses the \verb+get()+ function of Ducass\'e for this purpose also. By
using a very fast event comparison (as fast as 10ns to check for a
given source line, object, or method), it is possible to run the ODB
over a much wider range of programs.

\subsection{User Studies}

Several small-scale studies involving 2 - 8 subjects were
performed and gave encouraging results. An abstracted
version of an actual bug was presented to the subjects with
a brief explaination of the code and a half-hour
introduction to the ODB. The bug was a classic ``seen
snake-in-the-grass'' bug which took over an hour to find by the
original programmer using conventional tools.  All subjects
were able to pin-point the source within fifteen minutes.

\section{Performance and Memory Requirements}

The important thing to recognize here is that the ODB represents a
worst-case implementation--we can only do better. The ODB is completely
naive, it does no optimization at all. With aggressive use
of recomputation, collection can be reduced by orders of magnitude,
leaving us with insignificent slowdown and near-infinite recording space.
This allows us to focus on the main issue: Is this an effective way to debug a program?

On a 700MHz Apple 3G, the standard test method is one which takes two
arguments and performs four assignments to local variables. It requires
roughly 10$\mu$s to record the method call and another 1$\mu$s per each assignment
and marker event, for a total of 10 events in 19$\mu$s, and average of
2$\mu$s/event. 

Performance is not an issue for bugs which generate less than 10
million events.  At a rate of 2$\mu$s/event, it takes only 20 seconds
to fill the entire 2GB address space. 

For programs that depend on the GC, this is a issue. A couple typical 
programs are shown below.

\vspace{0,5cm}
\begin{tabular}{lr}
  Code&Slowdown\\
  Debugging ODB back-end&300x \\
  Debugging ODB display& 10x \\
  Debugging Ant& 7x

\end{tabular}
\vspace{0,5cm}

Notice that the back-end of the ODB is a worst-case program. It uses
no utility methods and manipulates few strings. Ant, by contrast, uses
a lot of utilities and libraries. The ODB display uses Swing extensively, but still
does a lot of array manipulations and \verb+JList+ construction.

A 64-bit address space changes things. It is then possible to collect
billions of events, and performance for data sets that fit becomes an
issue.  This implies that using \verb+get()+ to start and stop
recording will become more important.

In actual experience with the ODB, neither CPU overhead nor memory
requirements have proven to be a major stumbling block. Debugging the
debugger with itself is not a problem on a 110 MHz SS4 with 128 MB. On a 700
MHz iBook it's a pleasure. All bugs encountered while developing the ODB
fit easily into the 500k event limit of the small machine.

\subsection{Implementation}

The ODB keeps a single array of time stamps. Each time stamp is a 32-bit int
containing a thread index (8 bits), a source line index (20 bits), and a
type index (4 bits).  An event for changing an instance variable value
requires three words: a time stamp, the variable being changed, and the new
value. An event for a method call requires: a time stamp and a \verb+TraceLine+
object containing: the object, the method name, the arguments, and the
return value (or exception), along with a small pile of housekeeping
variables. This adds up to about 20 words. A matching \verb+ReturnLine+ will
also be generated upon return, costing another 10 words.

Every variable has a \verb+HistoryList+ object associated with it which is just a
list of time stamp/value pairs. When the ODB wants to know what the value of
a variable was at time 102, it just grabs the value at the closest previous
time. The \verb+HistoryLists+ for local variables and arguments hang off the
\verb+TraceLine+, those for instance variables hang off a ``shadow'' object which
resides in a hashtable. 

Every time an event occurs, it locks the entire debugger, a new time stamp
is generated and inserted into the list, and associated structures are
built. Return values and exceptions are back-patched into the appropriate
\verb+TraceLines+ as they are generated. 

The code insertion is very simple. The source byte code is scanned, and
instrumentation code is inserted before every assignment and around every
method call.

A typical insertion looks like this:

\begin{verbatim}
 289 aload 4                                  // new value
 291 astore_1                                 // Local var ie
 292 ldc_w #404 <String "Micro4:Micro4.java:64">
 295 ldc_w #416 <String "ie">
 298 aload_1                                  // ie
 299 aload_2                                  // parent Trace
 300 invokestatic #14 <change(String, String, Object, Trace)>
\end{verbatim}

where the original code was lines 289 and 291, assigning a value to the
local variable \verb+ie+.  The instrumentation creates an event that records
that on source line 64 of \verb+Micro.java+ (\#292), the local variable
\verb+ie+ (\#295), whose \verb+HistoryList+ can be found on the
\verb+TraceLine+ in register 2 (\#299), was assigned the value in register 1
(\#298).  The other kinds of events are similar.

\section{Integration with Event Analyzers}

The \verb+get()+ is a function which will match a modestly
complex pattern to an event. The pattern to be matched is based on prolog
syntax and proves to be very expressive and convenient. Because a moderately
complex query can be easily typed on a single line, it becomes possible to
implement \verb+get()+ as an extended minibuffer command. And this is a very good
thing.
 
The most common \verb+get()+ patterns will also map down to very efficient
code. A simple search pattern such as this:

\begin{verbatim}
port=enter & methodName="sort" & parmNames=["start", "end"]
\end{verbatim}

can run as fast as 10ns/test (7 cycles!). This pattern reads ``Find the entry
line in a method named \verb+"sort"+ which has two parameters, named \verb+"start"+ and
\verb+"end"+.'' Optimized code would require only six equality tests. The
less-than-optimal ODB adds about 50ns to this. Thus an \verb+get()+ test currently runs
about 40x faster than recording an event (50ns vs. 2us).

The conclusion is that using \verb+get()+ to start/stop recording can make a
huge impact on a long-running program, contrasted to running that same program
and relying on the garbage collector.

\section{Related Work}

There have been several previous attempts at doing something along
these lines. The most significant distinctions of the ODB are not related to
instrumentation and recording, but rather concern presentation and
navigation. The ODB is based on presenting \emph{state}, not events.
The ``debugger style'' interface is central to the ODB and navigation
is based on changing state (``What was the previous value of that
variable?''), rather than on the stream of events. The integration of
an event analyzer with a recording debugger is also unique.

In contrast, a number of projects have focused on ``reversible
execution'' and ``play back'' which record much the same information,
but don't allow the programmer the same view, nor the same navigation
techniques. Checkpointing and deterministic re-execution provide the
programmer the ability to use traditional breakpoints more quickly,
but are otherwise unrelated.

Starting in 1969, the EXDAMS project \cite{BALZER:1969} was a TTY-based system
which retained some amount of program state and allowed the programmer to
examine it post-facto. It is difficult to ascertain just how much it was
able to do. It was clearly hindered by the lack of a display.

Cohen and Carpenter \cite{COHEN:1976} built a system that collected and
stored trace information and allowed the user to make complex queries
about it. The interface was a Algol-like programming language for
making queries with TTY output.

``ZStep 95'' \cite{LIEBERMAN:1997} is a visualization/debugging tool for Lisp which
contains much of the same concept of retaining state and some of the
``navigation'' concepts. It has a significant emphasis on graphical
representations of call graphs and allows the programmer to ``play back'' the
execution of a program, video-style. 

Tolmach and Appel \cite{TOLMACH:1993} describe a debugger for ML which
is based on ``reversible execution.'' It has a similar concept in that
it allows the programmer to look at earlier state of a program based
on direct commands. It also relies on a TTY interface.

``Deja Vu'' \cite{ALPERN:2000} addresses the non-deterministic problems
of multithreaded programs by recording synchronization information,
thus allowing deterministic rerunning of the program. Ronsse
et. al. \cite{RONSSE:2000} also discuss the same problem.  They do not
attempt to use any of the information for debugging directly.

HERCULE \cite{RENAUD:2000} is a tool which can record and replay distributed
events, in particular, window events and appearence. It does for windows
much of what ODB does for programs, and provides much of the functionality
that the ODB lacks.

Trace analysis tools are generally designed as a sophisticated
breakpoint mechanism, sometimes accompanied by a query language
capable of deep program analysis \cite{DUCASSE:2000},
\cite{LENCEVICIUS:2003}.  These tools address a different problem than
Omniscient Debugging, and are relatively orthogonal. They are based on
a knowledgeble programmer making intelligent queries about suspicious
code. (The ODB allows the clueless programmer to wander around.) The
ODB incorporates the \verb+get()+ function of Ducass\'e for recording
control and dynamic queries.

\section{Conclusion}

For a large number of bugs, the ODB is highly effective. Problems that 
take hours with a breakpoint debugger can be solved in minutes. These are
the ``seen snake in the grass'' problems. The ``unseen snake in the grass''
problems are more challenging because the programmer has no obvious place to
start. Still, if the programmer has a reasonable idea of how the code works,
it is possible to search through the history to find a suspect value. Once
found, it is easy to verify.

For bugs that don't fit in the 10 million event limit of the 32-bit ODB, there are
a variety of approaches to make them fit. Garbage collection can be
effective for programs that generate an order of magnitude more
events. Beyond that, the run time can become excessive. A sophisticated
instrumentation tool, or a knowledgeble programmer, can specify methods that
don't need to be instrumented, radically reducing the number of events
recorded. An event analysis engine can select just the right moments for
turning recording on and off, also reducing the number of events recorded.

In practice, it has proven to be quite easy to make bugs fit into 10 million
events, and debugging even highly complex programs has not been a problem.
In addition to finding the bug, the ODB allows the programmer to confirm
exactly why it occured.

\bibliography{Bibliography}

\end{document}